\documentclass[structabstract]{aa}  
\usepackage{graphicx}
\usepackage{txfonts}
\usepackage{epstopdf}
\usepackage{longtable}
\usepackage{lscape}
\usepackage{natbib}
\usepackage{color}
\usepackage{siunitx}

\def\Msun{\hbox{$\thinspace M_{\odot}$}}
\def\Rsun{\hbox{$\thinspace R_{\odot}$}}
\def\Teff{\hbox{$\thinspace T_{\mathrm{eff}}$}}
\def\kms{\hbox{$\thinspace {\mathrm{km~s^{-1}}}$}}
\def\ms{\hbox{$\thinspace {\mathrm{m~s^{-1}}}$}}
\def\ALi{\hbox{$\thinspace A (\mathrm{Li})$}}
\def\vmic{\hbox{$\thinspace v_{\mathrm{mic}}$}}

\def\cc{\hbox{$\thinspace ^{12}\mathrm{C}/^{13}\mathrm{C}$}}

\bibpunct{(}{)}{;}{a}{}{,} 

\begin{document}

  \title{TAPAS - Tracking Advanced Planetary Systems with HARPS-N.  
  \thanks{Based on observations obtained with the Hobby-Eberly Telescope, which is a joint project of the University of Texas at Austin, the 
Pennsylvania State University, 
Stanford University, Ludwig-Maximilians-Universit\"at M\"unchen, and Georg-August-Universit\"at G\"ottingen.}
\thanks{Based on observations made with the Italian Telescopio Nazionale Galileo (TNG) operated on the island of La Palma by the Fundaci\'on 
Galileo Galilei of the INAF (Istituto Nazionale di Astrofisica) at the Spanish Observatorio del Roque de los Muchachos of the Instituto de Astrof\'{\i}
sica de Canarias.}
\thanks{Based on observations made with the Mercator Telescope, operated on the island of La Palma by the Flemish Community, at the Spanish Observatorio del Roque de los Muchachos of the Instituto de Astrofísica de Canarias.}
}

   \subtitle{II. Super Li-rich giant HD 107028.}

 \author{
          M. Adam\'ow,
          \inst{1,2}  
          A. Niedzielski,
          \inst{2}
          E. Villaver,
         \inst{3}
          A. Wolszczan,
	  \inst{4,5}
         K. Kowalik,
          \inst{6}                             	              
         G. Nowak,
          \inst{7,8,2}
          M. Adamczyk
          \inst{2}
     \and
         B.~Deka-Szymankiewicz
          \inst{2}
                   }

   \institute{ McDonald Observatory and Department of Astronomy, University of Texas at Austin, 2515 Speedway, Stop C1402, Austin, Texas, 78712-1206, USA.
    \email{madamow@astro.as.utexas.edu}
         \and
        Toru\'n Centre for Astronomy, Faculty of Physics, Astronomy and Informatics,
  Nicolaus Copernicus University, Grudziadzka 5, 87-100, 87-100 Toru\'n, Poland.
              \email{Andrzej.Niedzielski@umk.pl}
                                    \and
        Departamento de F\'{\i}sica Te\'orica, Universidad Aut\'onoma de Madrid, Cantoblanco 28049 Madrid, Spain.
         \email{Eva.Villaver@uam.es}
        \and
            Department of Astronomy and Astrophysics, Pennsylvania State University, 525 Davey Laboratory, University Park, PA 16802, USA.
         \email{alex@astro.psu.edu}
	 \and
             Center for Exoplanets and Habitable Worlds, Pennsylvania State University, 525 Davey Laboratory, University Park, PA 16802, USA.
             \and
             National Center for Supercomputing Applications, University of Illinois Urbana-Champaign, 1205 W Clark St, MC-257, Urbana, IL 61801, USA
           \and
Instituto de Astrof\'{\i}sica de Canarias, C/ V\'{\i}a L\'actea, s/n, E38205 - La Laguna,Tenerife, Spain.
 \and
Departamento de Astrof\'{\i}sica, Universidad de La Laguna, E-38206 La Laguna, Tenerife, Spain.  }

   \date{Received ; accepted }

 
  \abstract
{ Lithium rich giant stars are rare objects. For some of them, Li enrichment exceeds abundance of this
element found in solar system meteorites, 
suggesting
that these stars have gone through 
a Li 
enhancement  process.}
{We identified a Li rich giant HD~107028 with $\ALi>3.3$ in a sample of evolved stars observed within the PennState Toru\'n Planet Search. In this
work we study different enhancement scenarios and we try to identify  the one responsible for Li enrichment for HD~107028. }
{We collected high resolution spectra with three different instruments, covering different spectral ranges. We determine stellar 
parameters and  abundances of selected elements with both equivalent width measurements and analysis, and spectral synthesis. 
We also collected multi epoch high precision radial velocities in an attempt to detect 
a companion.}
{Collected data show that HD~107028 is a star at the base of Red Giant Branch. Except for high Li abundance,
we have not identified any other anomalies in its chemical composition, and there is no indication of a low mass 
 or stellar companion. We exclude Li production at the Luminosity Function Bump on RGB,
as the effective temperature and luminosity suggest that the evolutionary state is much earlier than RGB Bump.
We also cannot confirm the Li enhancement by contamination, as we do not observe any anomalies that are associated with this scenario.
}
  {After evaluating various scenarios of Li enhancement we conclude that the Li-overabundance of HD 107028 
  originates from Main Sequence evolution, and may be caused by diffusion process.}

\authorrunning{M. Adam\'ow et al.}
\titlerunning{TAPAS - II. HD~107028}
\maketitle
%

\section{Introduction}
 Lithium depletion is a natural prediction of stellar evolution. Not long after stars leave 
 Main Sequence they  experience mixing, called First Dredge Up (FDU) that will
lower the surface abundance levels to values of $\ALi = 1.5$\footnote{$\ALi=\log
\frac{n(\mathrm{Li})}{n(\mathrm{H})}+12$} dex or less. However, about a hundred
 Li-rich stars  ($\ALi > 1.5$) among Red Giant (RG) stars have recently been
found (\citealt{Kumar2011} and references therein,
\citealt{Ruchti2011,Lebzelter2012,MartellShetrone2013,Liu2014,Adamow2014}).

Lithium enhancement via internal production on Red Giant Branch (RGB)
can start at the earliest with Luminosity Function Bump (LFB), shortly after FDU ends.
LFB is a stage in RGB evolution for stars no more massive than $\approx 2.2\Msun$. 
It is associated with the removal of the molecular discontinuity that stems
from FDU processes. During LFB, star takes a small "step-back" in its evolution on
the HR diagram - effective temperature increases, while  luminosity decreases for a short time,
and stellar interior becomes turbulent \citep{Eggleton2006}.
Lithium may be produced during that evolutionary phase 
via a Cameron-Fowler mechanism \citep{CamFow1971} with the support of an
extra-mixing mechanism. 
Cameron-Fowler mechanism is probably responsible for Li production in stars on Asymptotic
Giant Branch (AGB). In this case, the thermal pulses can trigger the non-convective mixing.

As additional mixing on RGB begins
at the LFB, and the so-called Li-flash is also believed to be associated with this
phase \citep{CharBal2000,Pasquini2014}.
Thus, attempts to locate 
the Li-rich stars in the the HR diagram have driven a number of
studies trying to verify this Li-production mechanism (e.g., 
\citealt{Brown1989, Pilachowski1988, Pilachowski2000, Kumar2011, Ruchti2011,
Lebzelter2012, MartellShetrone2013}). However, these studies have shown, that Li-rich
stars can be found at different stages along the RGB evolution.
What is even more interesting, 
 one of the most Li rich star, SDSS~J0720+3036 with $\ALi=4.55\pm0.20$
\citep{MartellShetrone2013} was identified to be a subgiant.

All the above have led to proposals of a number of alternative mechanisms invoking
the presence of an extra-mixing process or pollution scenarios to explain 
the high photospheric values. Thus, it has been argued that lithium could
 be enhanced as a consequence of the
stellar readjustments following the accretion of substellar bodies \citep{Alexander1967,SiessLivio1999} or 
via the accretion of  dense interstellar medium clouds which chemical
composition was altered by a nearby
supernova explosions  \citep{WoosleyWeaver1995}. Lithium can also
be externally delivered by a companion in a binary system if one of the components has
already undergone lithium production on the AGB \citep{SacBoo1999}. All the scenarios based on external contamination,
involve anomalies in chemical abundances other than Li,
such as enhancement of neutron capture 
elements, that are produced via {\it s}-process in evolved stars, or $\alpha$ 
and {\it r}-process elements, associated with supernovae (i.e. \citealt{Terasawa2001}).
The accretion of substellar bodies should also increase stellar rotation \citep{Carlberg2010,Carlberg2012}
and may cause intense mass loss \citep{delaRezaDrake1997}
The latest searches for planets around giants have provided an opportunity to study the possible
connection between lithium abundance and hosting companions of different masses.
A planetary companion was found to the Li-rich red giant BD+48~740  \cite{Adamow2012} and 
in a recent study done for a sample of giants observed
within the PennState - Toru\'n Planet Search (PTPS) project, it is found that Li overabundance is
in many cases associated with a presence of companions \citep{Adamow2014}.

All scenarios of 
Li enhancement on RGB can potentially explain moderate
Li overabundances ($\ALi \lesssim 2.5$), but do not explain the existence of
extremely lithium rich stars.
About a hundred of lithium-rich giants  have been discover so far. About 14 
of there are so-called
super lithium-rich giant stars \citep{Brown1989, Balachandran2000, Reddy2005,
Carlberg2010, Kumar2011,Ruchti2011,MartellShetrone2013,Monaco2014}.
These evolved stars have abundances exceeding the interstellar
medium value, i.e. $\ALi=3.3$ \citep{Asplund2009}. 

In this paper we report a chemical abundance analysis of the Li-rich giant star
HD~107028. We found the star to have a $\ALi$ exceeding the value of 3.3 determined for meteorites.
HD~107028 has been as well the subject of a systematic search for a substellar companion
using HARPS-N at the 3.6~m Telescopio Nazionale Galileo under our program TAPAS
(Tracking Advanced Planetary Systems with HARPS-N)  after it was shown that  HET
observations show radial velocity (RV) variation of $\SI{36}{m.s^{-1}}$, significantly larger than expected 
stellar jitter. 

The paper is organized as follows: in Section
\ref{observations} we present the observations obtained for this target and we
outline the reduction procedures; 
Section \ref{stellar_params} presents stellar parameters;
In Section \ref{abundances} we present abundances determinations;
Analysis of available RVs is discussed in Section \ref{RV};
Section \ref{LPV} presents data on mass loss and activity;
and  finally Section \ref{discussion}, which includes
 the discussion and conclusions, closes the paper.
 
\section{Observations and data reduction. \label{observations}}

\begin{table*}
\centering 
\caption{Spectroscopic material used in this work.}
\begin{tabular}{lrrcc}
\hline \hline
Instrument & range & resolution &S/N &date\\
\hline
HET/HRS & 4076 -- \SI{7838}{\angstrom} & 60 000 &320 at \SI{5936}{\angstrom}&14 May 2006\\
HET/HRS & 7109 -- \SI{10917}{\angstrom}& 60 000 &171 at \SI{8991}{\angstrom} & 03 Mar 2013\\
TNG/HARPS-N & 3830 -- \SI{6930}{\angstrom} &115 000 & 141 at \SI{6582}{\angstrom} & 28 Apr 2013\\
Mercator/Hermes & 3770 -- \SI{9000}{\angstrom} & 85 000 & 178 at \SI{7124}{\angstrom} & 22 Jan 2015\\
\end{tabular} 
\label{spec_list} 
\end{table*}

HD 107028 (BD+69 657, HIP 59975) is a $V=7.66$, $B-V= +0.84$ and $U-B= +0.45$
\citep{Mermilliod1986} G5 \citep{HDCat} giant in Draco. New reduction of
Hipparcos parallaxes delivers $\pi=6.15 \pm 0.49$ and places the star at a
distance of 163~pc \citep{vanLeeuwen2007}. 157 epochs of Hipparcos 
 photometry reveal constant light of H$_p$=7.8035 $\pm$
0.0013 mag with a scatter of 0.015 mag.
The star belongs to  sample of evolved stars observed within
the PTPS project. Because of its
strong Li lines and the RV variations that could not be resolved with HET precision,
it was selected for TAPAS project.

The spectroscopic observations presented in this paper were made with three 
instruments (see Table \ref{spec_list}).The observation started with 
 the 9.2~m effective aperture  Hobby-Eberly Telescope (HET,
\citealt{Ramsey1998}) and its  High Resolution Spectrograph (HRS,
\citealt{Tull1998}) in the queue scheduled mode \citep{Shetrone2007}.
We also observed HD~107028 with  the 3.58~m
Telescopio Nazionale Galileo (TNG) and its High Accuracy Radial Velocity
Planet Searcher in the North hemisphere (HARPS-N, \citealt{Cosentino2012}) and
with 1.2~m Mercator telescope and Hermes spectrograph \citep{Raskin2011}.
The instrumental set-ups,
reduction and radial velocity measurements for HET/HRS and TNG/HARPS-N were
described in detail in \cite{tapas1} (hereafter Paper 1). We obtained altogether sixteen HET/HRS
spectra with typical signal-to-noise ratio (S/N) of 150-200 and 22
TNG/HARPS-N spectra with S/N of 30-100. We also obtained one HET/HRS spectrum
made in a different than standard PTPS spectrograph configuration. This
spectrum covers more red wavelength region and includes CN lines at
$\lambda\SI{8003}{\angstrom}$. We used standard IRAF scripts or HET/HRS data reduction.
 Data reduction for Mercator/Hermes spectrum
was done using the automated pipeline.

\section{Stellar parameters}\label{stellar_params}

For abundance analysis we used the four most suitable spectra available to us:  
two blue ones, with the highest S/N, one from
TNG/HARPS-N and the other from HRS/HRS (only in GC0 mode);  
and two red ones,  HET/HRS  spectrum that covers $\lambda\SI{8000}{\angstrom}$ region and
one Mercator/Hermes spectrum.

\begin{figure}
   \centering
   \includegraphics[width=0.5\textwidth]{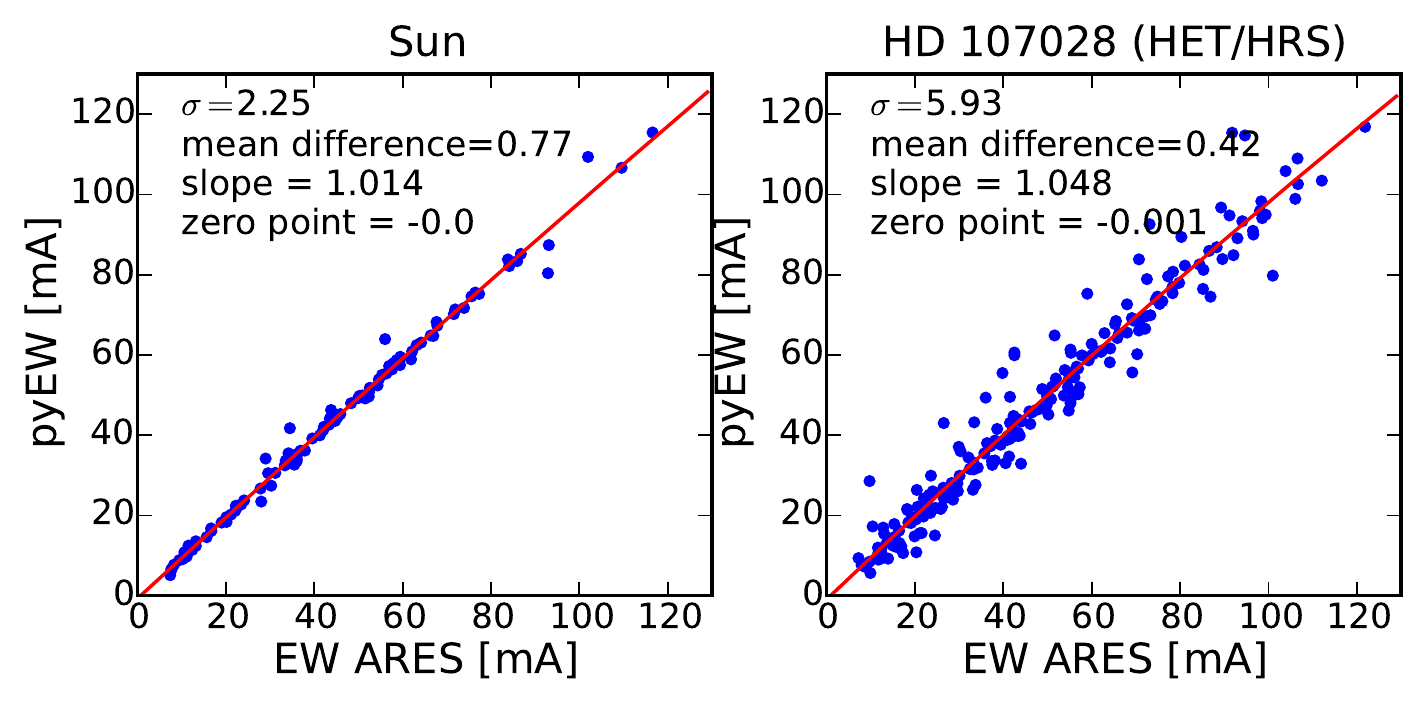}
    \caption{Comparison of equivalent widths determined with ARES and pyEW for the Sun and HD~107028.}
         \label{fig1}
         \end{figure}

The first estimate of stellar parameters was done using the available photometric data.
For this purpose we used BVJHK magnitudes from TYCHO and 2MASS catalogs and
$\Teff$ - color calibrations from \cite{RamirezMelendez2005}. 
From this analysis, the effective temperature of HD~107028 
was estimated as 5100~K but no reliable $\log g$ was available.
The luminosity of HD~107028 was calculated from the Hipparcos 
parallax \citep{vanLeeuwen2007}.
The effective temperatures agree within uncertainties.

To constrain the atmospheric parameters, mass and luminosity of HD 107028 as well as 
possible, we repeated the analysis with new spectra and a new method.
For this purpose, we developed a set of Python functions to 
measure equivalent widths (EWs) of  Fe~I and Fe~II lines 
and to evaluate stellar parameters using the current version of the MOOG\footnote{
http://www.as.utexas.edu/$\sim$chris/moog.html} stellar line analysis code \citep{Sneden1973}.

The pyEW\footnote{
https://github.com/madamow/pyEW}
 Python procedures for determination of EWs is based on ideas used in the ARES code \citep{Sousa2007}.
It redetermines continuum in a  small portion of the spectrum (usually \SI{2}{\angstrom}),
calculates derivatives to detect positions of lines and fits multiple gaussian profiles to the observed spectrum.
In Figure \ref{fig1} we present the comparison of equivalent widths determined with pyEW and
ARES for Sun (100 lines) using solar HARPS spectrum distributed with ARES code,
and for HD~107028 (208 lines).
For the latter we used the HET/HRS spectrum. In both cases we obtained very good agreement,
although the scatter, as expected, is higher for the HET/HRS spectrum given its lower 
S/N and resolution compared to the solar spectrum.

\begin{table*} 
\centering 
\caption{Summary of parameters for HD 107028 and Sun.}

\begin{tabular}{c|ccc||c} 
\hline
\multicolumn{4}{c}{Fe  lines by Takeda}\\
\hline Parameter & HET/HRS& HERMES &HARPS-N &Sun\\ 
\hline
$\Teff$        & $5124 \pm 41$ &	$5109\pm38$ 	& 	$5104\pm39$ &$5790\pm 47$	\\
$\log g$      & $2.97\pm0.06$ &	$3.01\pm0.06$ &	 $3.0\pm0.06$ &$4.44\pm0.06$ \\ 
$\vmic$      & $1.05\pm0.06$& 	$1.11\pm 0.05$& 	$1.03\pm 0.06$ & $0.97\pm0.06$\\
$[$Fe/H$]$ & $-0.25 \pm0.06$ &$-0.30\pm0.05$ & $-0.26 \pm 0.06$&$-0.09\pm0.06$\\
EP slope & 0.00238 &0.00706 & -0.00308&-0.01773\\
RW slope& 0.00121 &-0.00465 & -0.04525&0.02918\\
$n_{\mathrm{ Fe I}}$ &194 & 227& 193&173\\
$n_{\mathrm{ Fe II}}$ & 15 &18 & 14&14\\
\hline\hline
\multicolumn{4}{c}{a compilation of Fe and Ti lines with laboratory data}\\
\hline Parameter & HET/HRS& HERMES &HARPS-N&Sun\\ 
\hline 
$\Teff$        & $5190\pm 70$&$5139\pm67$ & $5132\pm64$&$5833\pm 63$\\
$\log g$      & $2.96\pm0.11$&$2.94\pm0.11$ & $2.94\pm0.11$&$4.37\pm0.13$\\
$\vmic$      &$1.19\pm0.10$& $1.07\pm0.09$& $1.04\pm 0.09$&$0.96\pm0.08$\\
$[$Fe/H$]$ &$-0.32\pm0.10$& $-0.31\pm0.10$&  $-0.23\pm 0.09$&$-0.03\pm0.08$\\
EP slope & 0.01685& -0.00414& 0.00675&-0.00641\\
RW slope& -0.025578& -0.00981& -0.02349&0.02659\\
$n_{\mathrm{ Fe I}}$ & 78& 103& 96&96\\
$n_{\mathrm{ Fe II}}$ & 33& 38& 31&33\\
\hline
\end{tabular} 
\label{params_results}
 \end{table*}

The analysis was done using  three different spectra (the red HET/HRS was not included in analysis)
and two line sets.
The first set of Fe I and II lines is adopted from \cite{Takeda2005b},
 who calibrated  line data on the solar spectrum. 
The second set of neutral and ionized Fe and Ti features is a compilation of lines for which $\log gf$
are determined in laboratory studies. The $\log gf$ values for Fe~I are taken from 
\cite{DenHartog2014,Ruffoni2014} and \cite{OBrian1991}; Fe~II lines are 
from \cite{MelendezBarbuy2009}. We also added Ti~I and Ti~II lines with $\log gf$ values taken from 
\cite{Lawler2013} and \cite{Wood2013}, respectively.
obtained by the University of Wisconsin atomic physics group.
From this compilation we chose isolated and moderately blended lines that are
present in our spectra. For the analysis we used only lines with equivalent widths
in the $\SI{5}{\milli\angstrom}<\mathrm{EW}<\SI{150}{\milli\angstrom}$ range.
We used only lines with $\lambda>\SI{4350}{\angstrom}$ due to the
 low S/N on the blue end of all spectra of HD 107028.

In determining atmospheric parameters we used MOOG's {\it abfind} driver
executed from a python wrapper
that iterates input stellar parameters and evaluates MOOG results. 
Iteration is made in the ($\Teff, \log g, \vmic$) space with the kMPFIT algorithm (a Python implementation 
of the Levenberg-Marquardt technique to solve the least-squares problem)
which is a part of the Kapteyn package\footnote{
https://www.astro.rug.nl/software/kapteyn/index.html}.
Adopted starting  parameters are: $\Teff=5100$ (obtained form photometric data), $\log g=3.0$,
$\vmic=1.0$ and [Fe/H]=0.0. For the calculations, we used  a grid of atmospheres by \cite{CastelliKurucz2004}.
For a given set of $\Teff, \log g, \vmic$ and [Fe/H],  we update the atmosphere model using a tool
by Andy McWilliam and Inese Ivans  for interpolating between grids.
During every iteration, all  abundances for single Fe~I and Fe~II lines 
that differ from the average
Fe~I and Fe~II values (respectively) by more than  $1\sigma$ are removed from the fit.
This criterion might be considered too strict, but initial $\sigma$ is calculated for all
abundances from all automatically measured lines. Usually,  there are some faulty EWs measurements among them
that need to be removed from further analysis. In case of HD~107028, $1\sigma$ threshold means
eliminating all values for which the difference between single abundance and average abundance is greater than 0.5~dex.
Input file where lines and their EWs are listed, is not edited, therefore this procedure is repeated for every iteration. 

The results obtained for different line lists and stellar spectra are summarized 
in Table \ref{params_results}.
Uncertainties presented there are estimated in a way 
proposed by \cite{GonzalezVanture1998}.
The uncertainties in $\Teff$ and $\vmic$ were estimated based on parameters of
abundances versus excitation potential and reduced EWs fits (respectively) for the best model.
Uncertainty in $\log g$
is a quadrature sum of the uncertainty in $\Teff$ and the scatter in Fe~II abundances,
and for uncertainty in [Fe/H] is a quadrature sum of the uncertainties in $\Teff, \log g, \vmic$
and the scatter in Fe~I abundances.

The final parameters adopted for further analysis are:
$\Teff=5133 \pm53$~K, $\log g=2.97\pm 0.09$, $\vmic=1.08\pm0.08 \kms$ 
and [Fe/H]$=-0.28\pm0.08$ (averaged values). 
The effective temperature we obtained is in agreement 
with the one based on intrinsic colors.
These parameters  locate  unambiguously HD 107028 in the subgiants/giants area in the HR diagram.

\begin{figure}
   \centering
   \includegraphics[width=0.5\textwidth]{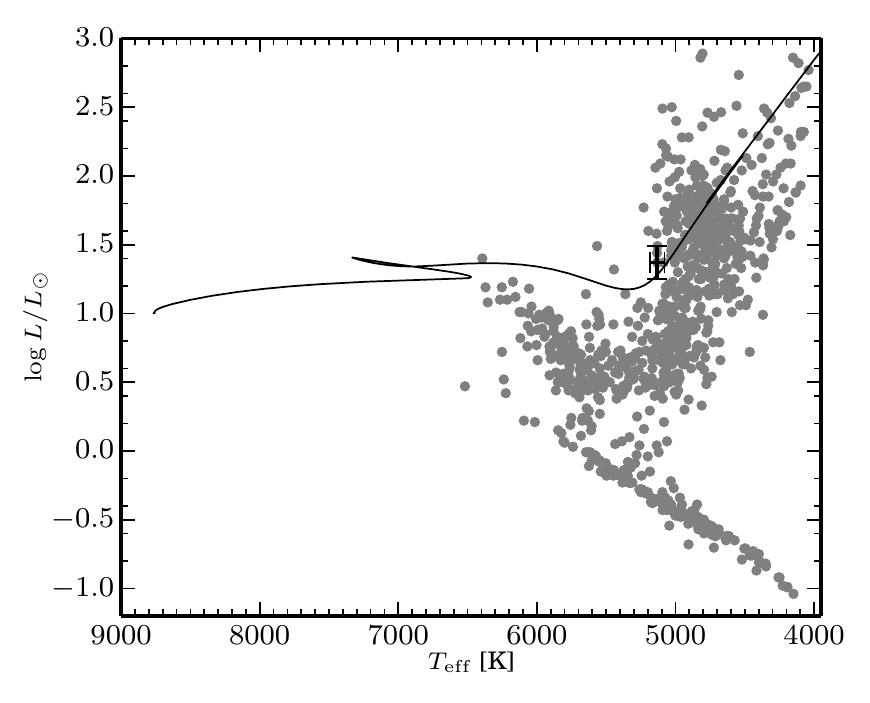}
      \caption{HD 107028 on top of all PTPS stars. Evolutionary track comes 
                    from PARSEC and it starts with hydrogen ignition.}
         \label{fig2}
         \end{figure}

The stellar mass and  age of HD 1070028 were estimated on the basis of spectroscopically determined
atmospheric parameters through construction of a probability distribution
function using an estimation by \cite{daSilva2006}  and
stellar isochrones from PARSEC (PAdova and TRieste Stellar Evolution Code,
\citealt{Bressan2012}). The stellar radius was estimated from the derived
masses and log g values obtained from the spectroscopic analysis, and using
the derived luminosities and effective temperatures. 
The adopted radius is the average of the two estimates. The $v_{rot}\sin i_{\star}$ 
was obtained via SME \citep{SME1996} by modeling a set of  spectral lines with a
procedure described by \cite{Adamow2014}. The stellar parameters  of
HD~107028 are summarized in Table \ref{Parameters} and its location 
on the Hertzsprung-Russell diagram is presented in Fig. \ref{fig2} .

\begin{table} 
\centering 
\caption{Summary of available data on HD 107028. }
\begin{tabular}{lll} 
\hline Parameter & value & reference\\ 
\hline\hline 
V [mag]& 7.57& ESA 1997 \\ 
B-V [mag] & 0.84 & \cite{Mermilliod1986} \\ 
$\pi$[mas]      &      $6.15\pm0.49$  & \cite{vanLeeuwen2007} \\ 
(B-V)$_0$ [mag] & 0.816 &  \\
M$_V$ [mag] & 1.54 &  \\
$\Teff$ [K] & $5133\pm53$ & This work  \\
$\log g$ & $2.97\pm0.09$&  This work\\
$[$Fe/H$]$ & $-0.28 \pm 0.08$& This work\\ 
$\vmic$ & $1.08\pm0.08$ & This work\\
RV [kms$^{-1}$] &-24.985$\pm$0.026& \cite{Nowak2012} \\
\hline \hline
$v_{\mathrm{rot}}\sin i_{\star}$ [$\kms$] & 1.46$\pm$1.19 & \cite{Adamow2014phd}\\
$M/\Msun$ & 1.72$\pm$0.21 & This work\\ 
$\log L/L_{\odot}$ &1.37 $\pm$0.12 & This work\\ 
$R/\Rsun$ & 6.6$\pm$1.0&This work\\ 
$\log \mathrm{age}$ & 9.20$\pm$0.14&This work\\ 
$V_{\mathrm{osc}}$ [ms$^{-1}$] & 3.01 & This work\\
$P_{\mathrm{osc}}$ [d] & 0.07 & \\ 
\hline 
\end{tabular} 
\label{Parameters} 
\end{table}

\section{Abundances \label{tab:abundances}} 
Lithium, CNO, neutron capture elements abundances and carbon isotopic ratio analysis was performed with a
python code that make use of the MOOG code ({\it synth} driver).
The use of spectral synthesis technique in obtaining abundances 
of those elements is motivated by the fact, that their lines 
are blended with other structures, might be affected by telluric lines or,
like in case of C and N, abundance analysis is based on molecular lines. 
The procedure
starts by building an initial synthetic spectrum, that then is cross correlated with the
observations to determine possible RV shifts. After that, an  automatic abundance 
analysis is performed, where fitting a requested set of chosen free parameters is
executed using the kMPFIT algorithm.  Except for abundances, 
we look for the best value of a parameter that describes gaussian smoothing of synthetic 
spectrum and we allow for small adjustments of the continuum level.
In this iterated procedure, new synthetic spectra with a
varying set of abundances is compared to the observed spectra in search for the
best fit.

We determined abundances of the $\alpha$ elements using the isolated lines 
selected by \cite{RamirezAllendePrieto2011}  from the measurement of their equivalent widths. 
Our final abundances for HD 107028 are summarized in Table \ref{tab:abundances}.

\subsection{Lithium}

\begin{figure}
   \centering
   \includegraphics[width=0.5\textwidth]{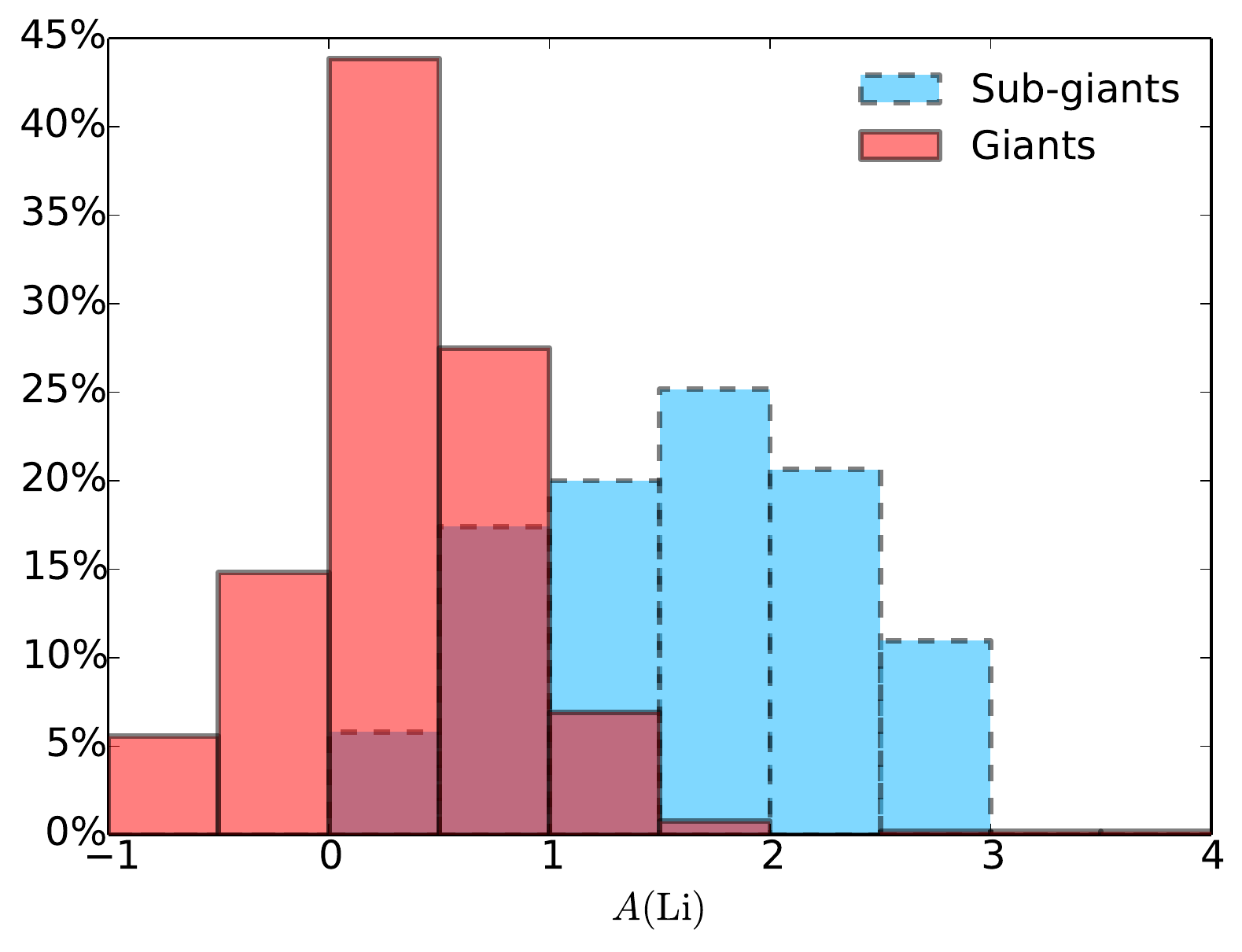}
    \caption{Distribution of Li abundances for evolved stars observed within PTPS \citep{Adamow2014phd}.}
         \label{fig3}
         \end{figure}

The Li I \SI{6708}{\angstrom} resonance feature is the one of the strongest lines in
HD~107028 spectra and proves that this star should be
classified as a Li-rich giant (see also Fig.~\ref{fig3} for lithium abundance distribution for evolved stars from PTPS sample).

\begin{figure}
   \centering
   \includegraphics[width=0.5\textwidth]{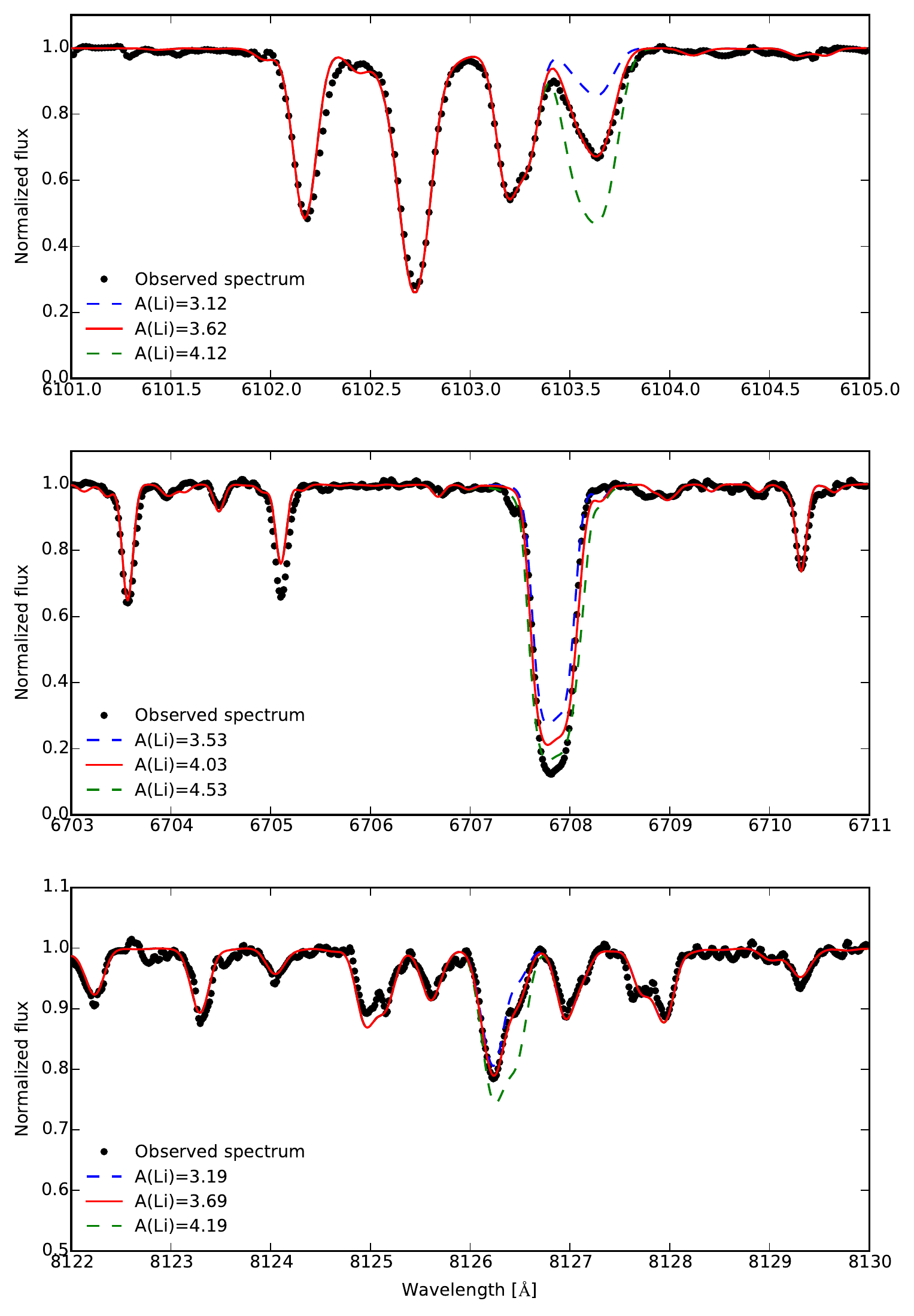}
      \caption{ Lithium lines identified in HD~107028 spectrum.}
         \label{fig4}
         \end{figure}

 The subordinate 
\SI{6104}{\angstrom} line is also detectable in this star, but it is much
weaker and located near strong Fe structures. Lithium abundance for HD~107028
was determined using those two lines from TNG/HARPS-N spectrum with the best
signal to noise ratio (141 around \SI{6708}{\angstrom} and 152 close to \SI{6103}{\angstrom}). 
The obtained values, presented in Table \ref{tab:abundances}, are in agreement, however the
abundance based on the stronger, redder line might be underestimated, as the model spectrum
does not reach the core of the spectral line (Fig. \ref{fig4}, middle panel). 
In the case of  HD~107028, the imperfect fit of
the bottom of the line may be caused by the fact that models might not
properly consider the environment where the \SI{6708}{\angstrom}~line is formed. 
This difference in abundances derived from the two Li lines is also
observed for other Li rich giants, i.e. for those discussed in
\cite{MartellShetrone2013}. Lithium lines may be subject
to several non-LTE processes (\citealt{Carlsson1994} and references therein). We
applied the non-LTE corrections  provided by \cite{Lind2009} to our LTE  lithium
abundances, calculated for a star of given  $\Teff,\; \log g,\; \vmic$ and [Fe/H].

The Mercator/Hermes spectrum includes also red Li lines at \SI{8126}{\angstrom}. Those lines
are blended with Fe and CN structures and are located in a region of the spectrum
that may be affected by H$_2$O telluric bands. Telluric correction has not been
applied to the Mercator/Hermes spectrum, but, based on the plot of the spectrum 
in Fig. \ref{fig4} (lower panel) where the best fit
of the stellar synthetic spectrum to the observational data is presented, there
is no significant telluric contamination close to the Li line that could
influence the abundance determination. The obtained value of $\ALi$ of 3.64 is in good agreement
with the value determined from the  $\lambda\SI{6103}{\angstrom}$ line.

\subsection{CNO  abundances and carbon isotopic ratio}

Oxygen abundance was obtained from the analysis of the  lines at $\lambda6300$,
and \SI{6363}{\angstrom}. Since these two lines are not affected by non-LTE effects,
no non-LTE correction was applied to determine the oxygen abundance.  
Both lines lay in a region affected by telluric lines, hence spectral analysis was performed
on a spectrum to which telluric division had been applied.
The line at $\lambda\SI{6300}{\angstrom}$ is also blended with a Ni line and lies in region where CN bands
may influence the analysis. 

The oxygen abundance was also obtained from the analysis of the $\lambda$7772, 7774,
and \SI{7775}{\angstrom} triplet, available from HET/HRS and Mercator/Hermes spectra. 
Since the oxygen triplet is strongly affected by non-LTE effects, we applied the empirical 
non-LTE corrections by \cite{AfsarSnedenFor2012}. 

Carbon abundance was determined by modeling of several C$_2$ (near \SI{5160}{\angstrom} and
\SI{5635}{\angstrom}) and CH bands (4230 - \SI{4240}{\angstrom} and 4300-\SI{4325}{\angstrom}). 
The abundance of carbon obtained is [C/H]$=-0.50\pm 0.05$.

The abundance of Nitrogen was determined based on CN bands near \SI{8003}{\angstrom},
using the red HET/HRS spectrum  
and the Mercator/Hermes spectrum. Although
the HET/HRS spectrum has a low signal to noise ratio, the values measured 
from both spectra are in agreement. Nitrogen, with an  abundance of [N/H]=0.6 is
enhanced in the HD~107028 atmosphere. However, it is important to
keep in mind that the N abundance determined using CN bands strongly depends on
the C abundance introduced in the analysis. 

The Mercator/Hermes spectrum, that targets CN bands at \SI{8003}{\angstrom}, was used to
determine the carbon isotopic ratio. The $^{13}$C lines are visible but they are not strong, and the obtained 
 $\cc$ ratio is 25.  
The S/N covering the CN bands close to \SI{8003}{\angstrom} is too low to determine reliable $\cc$
carbon isotopic ratio using the HET/HRS spectrum. 
Therefore, we used, for this purpose, the CH bands near 4230 and \SI{4370}{\angstrom} 
and we compared the observational data with a synthetic spectra using MOOG assuming different carbon isotopic ratios.
The determinations based on different molecules are in agreement, showing that in HD~107028 the FDU is in progress.
Line data for C$_2$ comes from \cite{Ram2014} and \cite{Brooke2013}, CN and CH data
were taken from \cite{Sneden2014} and \cite{PlezCohen2005}, respectively.

\begin{table} 
\centering 
\caption{Summary of abundances on HD 107028. Errors for $A(\mathrm{X})$
are standard deviations and they cover line to line deviations only.}
\begin{tabular}{llllrr} \hline \hline Element & $A(\mathrm{X})_{LTE}\pm\sigma$ 
& $A(\mathrm{X})_{NLTE}$& $A(\mathrm{X})_{\odot}$& $[$X/H$]$ & $[$X/Fe$]$ \\ 
\hline
Li$_{\SI{6104}{\angstrom}}$ & 3.62 &3.73&1.05 &&\\
Li$_{\SI{6708}{\angstrom}}$ & $4.03$&3.54&1.05 &&\\
Li$_{\SI{8126}{\angstrom}}$ & 3.69 &&1.05 & &\\ 
C  & $7.93\pm0.05$& & 8.43 & -0.50 &-0.23\\ 
N & 8.22  & &7.83 &0.39&0.66 \\
O&  $8.47\pm0.02$ &  &8.69 &-0.22  &0.05\\ 

 \hline\hline
 \multicolumn{6}{c}{$\alpha$ elements}\\
 \hline
 Mg &$7.55 \pm 0.15$&&7.60 & -0.05 &0.22\\
 Si &$7.36 \pm 0.06$&&7.51& -0.15 & 0.12\\
 Ca &$6.20\pm 0.07$&&6.34&-0.14 &0.13\\
 Ti &$4.75 \pm 0.09$&&4.95&-0.20 &0.07\\
 \hline \hline
 \multicolumn{6}{c}{Neutron-capture elements}\\
 \hline
Sr II & 2.49& &2.87& -0.38 & -0.11\\ 
Ba II  & 2.12 & & 2.18 & -0.06& 0.21\\
La II& $0.74\pm0.17$ & &1.10 & -0.36 &0.17\\
Eu II& 0.34 & & 0.52 &-0.18 & 0.09\\
 \hline \hline
 \multicolumn{6}{c}{Other elements}\\
 \hline
  Na &$6.19 \pm 0.05$ &&6.24 & -0.05 &0.22\\
  Al &$6.35 \pm 0.07$&&6.45& -0.10 & 0.17\\
 Ni &$6.15 \pm 0.08$&&6.22&-0.07 &0.20\\
\hline

\end{tabular} 
\label{abundances}
 \end{table}

\subsection{Abundances of $\alpha$-group, neutron capture elements and kinematics}

$\alpha$ element abundances were measured  using EWs for
several Na, Mg, Al, Ca, Ti and Ni lines selected from \cite{RamirezAllendePrieto2011} 
and using their line data. EWs were obtained automatically with pyEW
by fitting gaussian profiles to selected lines using the Mercator/HERMES spectrum.
Abundances were obtained using MOOG's {\it abfind} driver.

For the determination of  abundances of neutron capture elements we
used Ba II line at $\lambda \SI{6141}{\angstrom}$, two La II line at $\lambda 6262$ and \SI{6390}{\angstrom} and one Eu II line
at $\lambda \SI{6645}{\angstrom}$. All those lines have complex structures with isotopic and
hyperfine splitting, therefore we used spectral synthesis to
analyze them. Line data for those transitions were taken from \cite{Lawler2001a}
and  \cite{Lawler2001b}. 
Strontium abundance was calculated using line at $\lambda \SI{4077.71}{\angstrom}$.
All derived abundances for those elements are shown in Table \ref{tab:abundances}
and they are consistent with the abundances of nearby giants of similar metallicity 
(see i.e. \citealt{LuckHeiter2007} and \citealt{AfsarSnedenFor2012}).

The overall results of our abundance analysis are presented in Table \ref{tab:abundances}.  
We find for HD~107028 typical values of a thin-disk, sub-solar metallicity star and to check its consistency we
also computed space velocity components ($U, V, W$) following 
\cite{Johnson1987} but using the J2000 epoch in the coordinates and the
transformation matrix, and parallax and proper motions from the Tycho/Hipparcos
Catalog \citep{vanLeeuwen2007}. We obtain spatial velocities 
$(U,V,W)=(-87, -59,-51)\kms$ and the distance to the galactic plane $z=0.13$~kpc.
Hence, according to the \cite{Ibukiyama2002} criteria,
HD~107028 is a member of the galactic thin disk population,
This result agrees with the chemical composition of this star, typical for 
an object from the galactic thin disk.

\section{Radial velocities \label{RV}} 
HD 107028 was observed within PTPS  at 15 epochs over 2669 
days between  MJD=53755.47 and 56425.15. It showed low level RV 
variations of $36 \ms$ with average uncertainty of $6 \ms$ (i.e. $6 \sigma$). 
The RV variability was found to be over an order of magnitude larger 
than the expected amplitude of p-mode oscillations. 
We observed also bisector (BS) variations of $82 \ms$ with an average uncertainty of 
$16 \ms$ and of unclear nature but uncorrelated with the RV (r= -0.03).
No periodic signal was found in these data. A cross-correlation analysis 
of line profiles in the HET/HRS blue CCD GC0 spectrum of HD 107028 (unaffected by I$_2$ lines) 
as well as its GC1 spectra (cleaned from I$_2$ lines in the process of radial velocity measurements, 
\citealt{Nowak2012,Nowak2013}), showed regular shape of cross correlation 
function and no trace of a companion.

We  attempted to resolve this variability with the more precise and stable 
TNG/HARPS-N instrument given that following i.e. \cite{Adamow2014}, 
Li-overabundance in giants seems to be related to presence of  
companions. After 21 epochs of observations with TNG/HARPS-N taken over 517~days between 
MJD=56277.26 and 56795.04 we found RV variations of only $22 \ms$ with average 
uncertainty of $1.2 \ms$ and uncorrelated (r=-0.19) BS variations of only $10 \ms$.
Neither these data alone, nor combined with HET, showed a periodic signal. 
Our hybrid approach to model the data (see Paper 1 for details) also failed.

With RV measurements only, we cannot definitely
exclude that this star hosts a companion, as the orbital inclination of the potential
companions may be such that $\sin i \approx 0$. This kind of "pole on" systems are
not detectable for radial velocity searches.
However, the low level of observed RV variations, with  an amplitude of 
only $22\ms$, the lack of correlation between RV and BS 
and regular shape of the cross correlation function 
suggests better to  consider HD~107028 to be a single star.
The mean absolute RV of HD 107028 is $-24.485 \pm 0.007 \kms$. 

This radial velocity differs from an earlier measurement ($-29.20 \pm 1.2\kms$;
\citealt{Wilson1953}) by $\sim4.7\kms$.  This point cannot be included in RV fitting for several reasons:
there is no date and time when this RV measurement was done, the precision of measurements 
is very different and finally, time separation between Wilson's and our measurements is more than 60 years.
All that leads to unambiguous orbital solutions, but may suggest that HD~107028 has 
a low mass  companion on a highly elliptical, long period orbit. 

We cannot definitely exclude the possibility 
that  our observations cover only the flat part of RV curve during one orbital period.
However, assuming that the amplitude in RVs should be close to $5 \kms$ and that we have observed
only flat fragment of RV curve,  orbit of companion hosted by HD~107028 should have $e\gtrsim0.8$
and  a very specific orientation in space (longitude of ascending node of $0^{\circ}$ or $180^{\circ}$). In any other cases,
a longterm trend in RVs should have been easily detected in TNG/HARPS or HET/HRS data, but we have not observed it.

Therefore, collected observations cannot definitely exclude the hypothesis that HD~107028 hosts 
a companion on an elliptical orbit and one needs to keep in mind that RV technique effectiveness 
depends on system orientation in space. However, we claim that the presence of such a companion 
is unlikely, as we do not see any long term trends in collected RV data. 
We also argue  that radial velocity measurement by \cite{Wilson1953} might be overestimated for HD 107028.

\section{Line profile variations: activity and traces of mass-loss \label{LPV}}

The Ca II H \& K line profiles are widely accepted as stellar activity
indicators. Both the Ca~II H \& K lines and the infrared Ca~II triplet lines at
$8498-\SI{8542}{\angstrom}$ lay outside HET/HRS wavelength range but the Ca II H \& K lines
are available to us in the TNG HARPS-N spectra and the 8498-\SI{8542}{\angstrom} ~range is present
in Mercator/Hermes spectrum. Although the S/N in the blue spectrum for this red
giant is not high, no trace of reversal, typical for active stars
\citep{EberhardSchwarzschild1913} is present, and no obvious chromospheric
activity can be demonstrated neither for TNG/HARPS-N nor for Mercator/Hermes
spectra. To quantify possible activity-induced line profile variations we
calculated the $S^{\mathrm inst}_{\mathrm HK}$ index according to the \cite{Duncan1991} 
prescription for all 21 epochs of HARPS-N observations. The mean value is
$S_{\mathrm inst}=0.35 \pm 0.02$ which places HD~107028 among moderately active
subgiants according to \cite{IsaacsonFischer2010}.

One of the explanations for Li overabundances in evolved stars implies that Li enhancement is
associated with mass ejections (in a form of dust) from the star (see e.g.
\citealt{delaRezaDrake1997, SiessLivio1999,Kumar2015}) that we would see as infrared excesses.
To verify whether  HD~107028 presents an infrared excess, we queried the
WISE \citep{WISE2010} and 2MASS catalogs. The brightness of the star in
K-WISE[12$\mu$m] and J-K 2MASS colors seems to be rather typical when 
compared to larger samples, like those presented by \cite{Lebzelter2012} or
\cite{Adamow2014}. Hence, we cannot report a significant infrared excess for
HD~107028.

If the star has an intensive mass-loss in the form of gas, it should
be reflected in the spectral line shape. \citep{Reddy2002} has shown that this
phenomena might be observed through the Na D lines profiles - an additional
component to sodium doublet should be present in the blue wing of stellar sodium
lines. Again, we have not observed any irregularities of Na D lines or
additional components to them. Therefore, HD~107028 does not seem to have 
gone recently through an intensive mass loss episode, neither in form of dust nor gas.

\section{Discussion}\label{discussion}

HD~107028 is a giant of $1.7\Msun$ that has recently started its evolution on  RGB.
Its combination of effective temperature,surface gravity and $\cc$ is consistent 
with a star that has already started the FDU and
passed the point where Li depletion starts and carbon isotopic ratio drops \citep{CharBal2000}.
Our derived
$\cc\sim 25$ value is in agreement with the value obtained by \cite{Sweigart1989}
for a star with $\Teff, \log L/L_{\odot}$ and [Fe/H] similar to those determined for HD~107028.

The prominent lithium overabundance for HD~107028, $\ALi\geq 3.6$,
places the star among the most Li-rich stars known and cannot be
explained with Li production mechanisms at the LFB, because this star has not reach this
phase yet, even if we assume the low mass star evolution model
proposed by \cite{Denissenkov2012}, where LFB occurs
for lower effective temperatures ($\sim 5000$~K).

Stars with high Li abundance that are still undergoing FDU can be misidentified 
as Li rich giants. These are intermediate mass stars that leave the MS with 
relatively high Li abundance 
and at the bottom of the RGB they may still have $\ALi\approx 1.5$, 
which is usually adopted as a low abundance limit for considering a star Li rich.
The lithium abundance observed in HD~107028 is higher than the one
expected for population I stars  on the MS ($\ALi=3.3$), 
even though this star has partially gone through the FDU process.  
Therefore, the chemical composition of HD~107028 must have been modified
by some Li enhancement process.

The first dredge up is the only known process that can alter the chemical
composition of single stars during the subgiant evolution, but it does not result in Li
(or any other element) production. Hence, the observed high Li abundance in HD~107028 might have
originated during the MS evolution of the star.
There are several other stars that reveal high Li abundance and 
have not reach RGB yet.
Three of them are turn-off stars located in clusters \citep{Deliyannis2002,Koch2012,Monaco2012}
and the forth one is a field subgiant detected by \cite{MartellShetrone2013}.

External pollution is an alternative to lithium nucleosynthesis in the stellar
interior, and it is expected to be associated with enhancements in the $\alpha$ group 
and neutron-capture elements. This, however, does not seem to be the case in HD~107028 since the Li
overabundance is not associated to $\alpha$ group, La or Eu enrichment, having all
typical values of a thin disc star with metallicity of -0.3.
 
Exoplanets observations show a lack of close planets 
($a<0.6$~AU) around evolved stars for which planetary engulfment is a very
plausible explanation (see e.g. \citealt{VillaverLivio2009, Villaver2014}). 
In principle, HD~107028 is evolved enough to have been able 
to ingest a hypothetical inner companion, given its orbit was close enough, 
like in known hot Jupiter systems, for which semi major axis is usually smaller than 0.05~AU
(stellar radius for HD~107028 is $6.6\Rsun \approx 0.03$~AU).
 
However, we find very unlikely that planet  engulfment has a strong influence on the Li enrichment 
found in HD~107028. First off all,
the HD~107028 Li abundance is too high. In order to raise the stellar Li abundance to a $\ALi=3.6$ level, 
 an unrealistic number of planets needs to be engulfed
(few hundreds Jupiter like objects). Their chemical composition must have  been
peculiar, i.e. their lithium abundance must exceed the one observed on the
interstellar medium (assuming the model for Li enhancement via engulfment
discussed by \citealt{Carlberg2012}).  Moreover, the presence of an external companion as the cause 
(or effect via dynamical interactions 
within a system that might have caused ingestion) of the anomalous Li measurement is unlikely as
precise radial velocities measurements obtained within the PTPS and TAPAS
projects  do not show any periodic signal that might be interpreted as a low mass 
companion.  We do not find any other signs of phenomena that could be associated with an
engulfment episode. HD~107028 is a slow rotator with $v \sin i = 1.5 \kms$, 
which is typical for an evolved star,  but on the other hand consistent with 
a pole-on rotating star. It also shows no signs of enhanced mass
loss, neither in a form of gas nor dust.

Contamination from more evolved stellar companions was suggested as 
the most likely explanation for the high Li abundance measured for the 
Li rich objects located in the NGC~6397 \citep{Koch2012} and M4 \citep{Monaco2012} clusters. 
This scenario cannot be applied to account for the Li abundance found in HD~107028 
as we do not find any signs of companions in the radial velocity measurements.  

\cite{Deliyannis2002}  proposed diffusion as
a process leading to Li overabundance in the turn-off star J37 in the NGC~6633 cluster. 
For this star Li overabundance is not the 
only observed anomaly in the chemical composition. Even though kinematic 
data and photometry locate J37 as a member of NGC~6633, its Fe abundance
is significantly (2-3 times) higher than for the rest of stars assigned to the cluster. 
Elements like Ni, S, Si, Ca and Al are also enriched, while carbon is depleted. 
Diffusion is a process that can explain a so-called Li-dip (or Li-gap) a sudden 
drop in Li abundances for MS stars in a specified effective temperature range.
But on the hot side of Li-dip, diffusion model predicts Li enhancement.
In this narrow range of effective temperature, Li abundance 
can be raised to the level of $\ALi \approx 4.7$ 
(see Fig. 2 in \citealt{Deliyannis2002}).

If we follow back the evolution of HD~107028 as a $1.7\Msun$ star to the MS  
(note we do not expect intensive mass loss between the MS
and its current location on the HR diagram as we do not see any signs of it in the data), 
we find its effective temperature on the MS was in the range 
from $\sim\SI{8900}{\kelvin}$ to $\sim\SI{6500}{\kelvin}$ (see Fig. \ref{fig2}).
 It means, that HD~107028 can fit into the diffusion model presented in 
 \cite{Deliyannis2002}. It is hard to say if the chemical composition of HD~107028  
differs from other stars (with the exception of Li) as this kind of comparison is more difficult 
for a field object from the galactic disk than for a member of stellar cluster.
Nevertheless, based on [Fe/H] vs. [X/H] trends  for stars in galactic disc, HD~107028
should be considered to be a rather typical star more than an outsider.
 HD~107028 chemical composition was altered by some extraordinary event or process, 
and for some reasons, 
this unknown process does not affects other elements as remarkably as it does with Li. 

In this work we have shown, that the nature of Li enhancement in HD~107028
is not connected to Li production on RGB or contamination scenarios. However, this object
is a very interesting case for studying Li processing in  stars.

\begin{table} 
\centering 
\caption{HET \& HRS measurements of  HD~107028. Radial velocity (RV), bisector
(BS) and their respective errors are measured in \si{\meter\per\second}} 

\begin{tabular}{ccccc} 
\hline MJD& RV & $\sigma_{\textrm{RV}}$ & BS & $\sigma_{\textrm{BS}}$\\
\hline\hline 
53755.475041 & -18.19 & 6.53 & -4.74 & 17.59 \\ 
53808.329433 &-10.37 & 4.84 & 24.00 & 16.14 \\
53843.191794 & -2.36 & 4.35 & 63.56 & 10.59 \\
53893.124855 & -8.02 & 6.54 & -18.12 & 11.80 \\ 
54463.522598 & -14.36 & 6.64 &7.16 & 21.12 \\ 
54520.372234 & -12.07 & 6.02 & 49.32 & 9.85 \\ 
56288.509421 &-2.34 & 6.88 & 45.15 & 22.29 \\ 
56308.463090 & -7.34 & 5.80 & 26.25 & 15.41 \\
56323.433646 & 8.92 & 5.57 & -1.68 & 14.49 \\ 
56335.412303 & 13.11 & 6.21 &47.18 & 19.64 \\ 
56354.352824 & 9.55 & 5.64 & 0.42 & 17.89 \\ 
56367.318287 &17.31 & 6.30 & -10.13 & 17.01 \\ 
56379.282778 & 9.95 & 5.29 & 4.65 & 15.76 \\
56395.245718 & 9.07 & 5.77 & 33.99 & 16.49 \\ 
56425.153495 & 4.00 & 5.89 & 43.69& 15.70 \\ 
\hline 
\end{tabular} 
\label{HETdata} 
\end{table}

\begin{table} 
\centering 
\caption{TNG \& HARPS-N measurements of HD~107028. Radial velocity (RV), its error and
bisector (BS)  are measured in \si{\meter\per\second}}.

\begin{tabular}{cccc} 
\hline 
MJD & RV &$\sigma_{\textrm{RV}}$& BS 
\\ \hline\hline
56277.260735 & -24495.51 & 0.81 & 35.19 \\ 
56294.242058 & -24488.44 & 0.71 & 35.85 \\ 
56321.142563 & -24481.60 & 1.00 & 42.60 \\
56374.005509 & -24493.47 & 0.74 & 33.45 \\ 
56410.913816 & -24475.76 & 0.63 & 36.11 \\ 
56411.009163 & -24492.93 & 0.68 & 33.82 \\
56430.967926 & -24486.64 & 0.83 & 38.61 \\ 
56469.876485 & -24490.62 & 1.31 & 38.47 \\ 
56469.959043 & -24490.51 & 1.59 & 32.90 \\
56647.252802 & -24492.09 & 1.83 & 41.94 \\ 
56685.115550 & -24475.23 & 1.23 & 32.95 \\ 
56685.213074 & -24473.66 & 1.30 & 33.72 \\
56685.275789 & -24475.27 & 1.61 & 32.89 \\ 
56739.967323 & -24488.57 & 1.08 & 42.47 \\ 
56740.025788 & -24491.21 & 1.27 & 39.73 \\
56740.101594 & -24486.80 & 1.63 & 41.03 \\ 
56740.142011 & -24488.17 & 1.15 & 41.92 \\ 
56770.022102 & -24487.05 & 1.02 & 38.44 \\
56770.070215 & -24486.01 & 0.96 & 36.24 \\ 
56794.901061 & -24473.29 & 1.29 & 38.94 \\ 
56795.040834 & -24480.07 & 1.92 & 35.49 \\
\hline \end{tabular} 
\label{HARPSdata} 
\end{table}

\begin{acknowledgements}
We thank the HET  and IAC resident astronomers and telescope operators  for
support.
We would like to thank dr~Sergio Simon Diaz from IAC for obtaining
the Mercator/Hermes spectrum used in this work. We also thank  dr Christopher Sneden
for comments that greatly improved the manuscript.
MA acknowledges the Mobility+III” fellowship from the Polish Ministry of Science
and Higher Education. 
MA, AN, BD and MiA  were supported by the Polish National Science Centre
grant no. UMO-2012/07/B/ST9/04415.  
E. V. acknowledges support from the Spanish Ministerio de Econom\'ia y Competitivid
ad under grant AYA2013-45347P.
KK was funded in part by the Gordon and Betty Moore Foundation's
Data-Driven Discovery Initiative through Grant GBMF4561.
This research was supported in part by PL-Grid Infrastructure.
The HET is a
joint project of the University of Texas at Austin, the Pennsylvania State
University, Stanford University, Ludwig- Maximilians-Universit\"at M\"unchen,
and Georg-August-Universit\"at G\"ottingen. The HET is named in honor of its
principal benefactors, William P. Hobby and Robert E. Eberly. 
The Center for Exoplanets and Habitable Worlds is supported by the Pennsylvania State
University, the Eberly College of Science, and the Pennsylvania Space Grant
Consortium.
This work is based on observations obtained with the HERMES spectrograph, 
which is supported by the Fund for Scientific Research of Flanders (FWO), Belgium, 
the Research Council of K.U.Leuven, Belgium, the Fonds National de la Recherche 
Scientifique (F.R.S.-FNRS), Belgium, the Royal Observatory of Belgium, 
the Observatoire de Gen\`eve, Switzerland and the Th\"uringer 
Landessternwarte Tautenburg, Germany.
This work makes use of Astropy, a community-developed core Python package for Astronomy,
 as well as SciPy and NumPy.

\end{acknowledgements}

\bibliographystyle{aa} 
\bibliography{tapas2.bib} 

\end{document}